\documentclass[manuscript, screen, nonacm]{acmart}
\usepackage{threeparttable} 
\usepackage{multirow}
\usepackage{subcaption}
\usepackage{caption}
\AtBeginDocument{%
  }

\begin{document}

\title{Change is Hard: Consistent Player Behavior Across Games with Conflicting Incentives}

\author{Emily Chen}
\affiliation{%
  \institution{Department of Computer Science, University of Southern California}
  \city{Los Angeles}
  \state{California}
  \country{USA}
}
\affiliation{%
  \institution{Department of Computational Media, University of California at Santa Cruz}
  \city{Santa Cruz}
  \state{California}
  \country{USA}
}

\author{Alexander J Bisberg}
\affiliation{%
  \institution{Department of Computer Science, University of Southern California}
  \city{Los Angeles}
  \state{California}
  \country{USA}
}
\affiliation{%
  \institution{Department of Computer Science \& Software Engineering, California Polytechnic State University}
  \city{San Luis Obispo}
  \state{California}
  \country{USA}
}

\author{Dmitri Williams}
\affiliation{%
  \institution{Annenberg School of Communication, University of Southern California}
  \city{Los Angeles}
  \state{California}
  \country{USA}
}
\author{Magy Seif El-Nasr}
\affiliation{%
  \institution{Department of Computational Media, University of California at Santa Cruz}
  \city{Santa Cruz}
  \state{California}
  \country{USA}
}

\author{Emilio Ferrara}
\affiliation{%
  \institution{Department of Computer Science, University of Southern California}
  \city{Los Angeles}
  \state{California}
  \country{USA}
}

\renewcommand{\shortauthors}{Chen et al.}
\makeatletter
\let\@authorsaddresses\@empty
\makeatother
\begin{abstract}
    This paper examines how player flexibility -- a player's willingness to engage in a breadth of options or specialize -- manifests across two gaming environments: League of Legends (League) and Teamfight Tactics (TFT). 
    We analyze the gameplay decisions of 4,830 players who have played at least 50 competitive games in both titles and explore cross-game dynamics of behavior retention and consistency. 
    Our work introduces a novel cross-game analysis that tracks the same players' behavior across two different environments, reducing self-selection bias. 
    Our findings reveal that while games incentivize different behaviors (specialization in League versus flexibility in TFT) for performance-based success, players exhibit consistent behavior across platforms.
    This study contributes to long-standing debate about agency versus structure, showing individual agency may be more predictive of cross-platform behavior than game-imposed structure in competitive settings. 
    These insights offer implications for game developers, designers and researchers interested in building systems to promote behavior change. 
\end{abstract}

\maketitle
\section{Introduction}

Understanding how users navigate multiple systems with different design constraints is fundamental to human-computer interaction. 
As users increasingly engage across different platforms, context switching from social media networks to entertainment systems to productivity tools, we face the central question of whether system design constraints shape user behavior or if users bring stable behavioral preferences that persist regardless of structural differences. 
This has profound implications for cross-platform design, system migration, user modeling and any context where users must adapt to different interface paradigms or incentive structures. 

Games provide a unique opportunity to study this question systematically.
Compared to most digital platforms, online games are simplified complex systems where behavior can be measured against clear structural constraints and explicit performance metrics, while also generating rich telemetry from large user populations. 
Competitive games add an additional dimension: because players have explicit performance goals, we can observe how they balance their intrinsic behavioral preferences against extrinsic pressures to optimize for success. 
Examining player behavior across games with different design philosophies allows us to empirically test how structural incentives interact with individual agency -- insights that extend beyond games to inform our understanding of user behavior across digital systems. 

This paper examines how players navigate two games with fundamentally different structural incentives for performance-based success: League of Legends (League) and Teamfight Tactics (TFT). 
League is a team-based multiplayer online battle arena (MOBA) that requires players to commit to a champion selection before the match starts.
This pre-game selection draft, along with the mechanical mastery required for individual champions and team-based nature of the game, creates incentives for specialization: players who develop an expertise in fewer champions tend to perform better. 
TFT's main game mode is an individual auto-battler that incorporates intentional randomness through unit availability and pushes players to dynamically adapt their strategies during gameplay based on what units appear and what opponents are building. 
This design creates incentives for flexibility, as players who can pivot between different compositions tend to perform better. 

Both of these games are developed by Riot Games and thus share the same ecosystem. 
By tracking the same 4,830 players across both League and TFT over the course of more than a year, we reduce self-selection bias that typically confounds cross-platform behavioral research; this also allows us to empirically test whether structural design successfully shapes behavior or whether individual dispositional preferences persist across game contexts. 
Research on personality stability suggests that core behavioral tendencies remain remarkably stable across contexts. 
While specific behaviors may vary, individuals maintain characteristic patterns in how they respond to different situations~\cite{mischel1995cognitive,fleeson2001toward,roberts2000rank}. 
This predicts that players' dispositional preferences for variety-seeking versus specialization should persist across games as stable individual differences, even when games reward different behaviors. 
However, the pursuit of success may drive some adaptation, particularly among elite players for whom performance is paramount. 
Our study design allows us to empirically test these competing predictions.
We focus on a user's flexibility, operationalized as the breadth of play styles (in League) or board compositions (in TFT) that players engage with over time, measured through end-game outcomes extracted from gameplay telemetry. 
A user's initial competitive rank at the time of data collection serves as a proxy for performance-seeking motivation, allowing us to examine whether highly motivated users show different patterns of behavioral adaptation to structural incentives. 

We approach this problem primarily through the lens of Giddens' Structuration Theory, which provides a framework for understanding how individual agency and structural constraints mutually constitute behavior~\cite{Giddens1979agency,stones2017structuration}. 
Individual motivations and preferences do not exist in a vacuum; users make decisions within systems that align with incentives and can also push against these same systems. 
The tension between our individual choices and the way systems act on us is at the heart of the classic sociology debate between agency versus structure. 
Previous work suggests that behavior is influenced by either structure -- external systems and rules -- or by agency -- individual thought and actions~\cite{tan2011understanding}. 
Giddens' Structuration Theory argues that both agency and structure play a role in shaping behavior and that the two are not mutually exclusive~\cite{stones2017structuration,Giddens1979agency,karp1986agency}.
Scholars across multiple disciplines studying this agency versus structure debate have pointed to the complexity of disentangling the two in empirical research; environment mechanics, social norms and system architectures vary across platforms and modalities, and can also evolve over time.
While much of the existing literature focuses on the challenges in translating findings between virtual and real-world behavior, these concerns are equally relevant in virtual-to-virtual comparisons, where affordance differences and participant awareness can complicate clean empirical tests~\cite{malaby_control_2006,williams2010mapping,baskan_social_2024,pozzebon_challenges_2005}.

Competitive gaming environments offer a unique lens to address this challenge and study the interplay between agency and structure. 
Games, such as League and TFT, provide rich contexts where a game's design -- its rules, incentives and constraints --  may incent or force players to make certain choices. 
This is the basis of Gibson's Affordance Theory~\cite{gibson2014theory}. 
One such choice in games, where a player can take on different roles each session, is a player’s preference to specialize or diversify their roles or play styles. 
In games where players are competing and trying to win, these choices are often influenced by whether the player perceives certain options to be more competitive.
Competition also means that players typically have a common goal that they are working towards: performance-based success. 
A shared objective enables us to investigate how behavior adapts in pursuit of optimal outcomes. 

To develop our hypotheses about the interplay between agency and structure, we draw on complementary theoretical frameworks.
Self-Determination Theory (SDT)~\cite{ryan_self-determination_2000} helps explain the motivational forces at play in pursuit of competitive success: the \textit{competence} motivation, which reflects users' sense of effectiveness and mastery, manifests differently across gaming contexts; depending on a game's structure and goals, users may need to adapt their play styles for optimal performance~\cite{valls-serrano_cognitive_2022,chen2024flexibility}.
Research on personality stability, however, suggests that dispositional preferences for flexibility versus specialization should persist across contexts as stable individual differences~\cite{fleeson2001toward,epstein1979stability}. 
Structuration Theory synthesizes these perspectives through its concept of the duality of structure.
Structures constrain behavior while individual agency can reproduce or transform those structures, with the balance depending on reflexive monitoring (conscious evaluation of practices) versus routinized behavior (automatic enactment of established preferences).

This paper examines a user's \textit{functional play style flexibility}, defined as the breadth or variability in a player's role or play style selection, and how this adaptability varies across different gaming environments.
This addresses a fundamental design question in games and more broadly in HCI: how do system constraints interact with user agency and autonomy, and can individual behavior preferences persist across systems with conflicting incentive structures? 
We do not assume players will or will not change their flexibility in response to structural differences, but empirically test whether behavioral patterns persist or adapt. 
Our study design, which holds individuals constant while varying game structure, allows us to observe these dynamics without presupposing the outcome.
This also allows our work to contribute to the agency-structure debate by being able to analyze the different impacts of both major factors, avoiding the issue of selection bias that faces most such work.

To answer this question, we develop an operationalization of flexibility that captures the breadth of play style choices within each game's unique design constraints, enabling direct comparison across different gaming environments.
We test three hypotheses about how individual agency and structural incentives interact to shape player behavior: (H1) game structure creates opposing relationships between flexibility and competitive success, driven by the competence motivation, (H2) individual dispositional preferences persist across games despite opposing incentives, reflecting stable personality traits and (H3) highly motivated players show partial adaptation through reflexive monitoring while maintaining baseline dispositional tendencies, demonstrating the duality of structure and agency.

Beyond the specific context of League and TFT, our findings offer insights into player identity and behavioral persistence more broadly, demonstrating that individuals maintain relatively stable behavioral dispositions even when environmental incentives actively reward different approaches.
We conceptualize flexibility preferences as a stable component of player identity that persists across gaming contexts, revealing the limits of structural design in influencing user behavior. 
This has implications for understanding how users navigate cross-platform experiences, resist or adapt to system design changes, and maintain consistent preferences across digital environments with different incentive structures.
Our findings also offer practical insights for both HCI researchers and game designers into how structural incentives interact with individual preferences, highlighting the role of flexibility as a measurable proxy for user autonomy. 
These insights can inform and are central to the design of more user-centered, adaptive and personalized cross-platform experiences.

\section{Background}
Games are virtual environments in which individuals make decisions about interacting with their surroundings and, in multiplayer contexts, with other individuals. 
Although behavior in games can differ from real-world actions -- and nuances of the gaming community must be considered -- studies of behavior in games can offer valuable insight into broader behavioral preferences~\cite{williams2010mapping}.
Virtual worlds can also provide a window into decision-making in online environments, which is becoming an increasingly important area of study as more of our lives migrate to digital spaces. 

\subsection{Player Behavior in Online Multiplayer Games}
Research on player behavior in online games has evolved from early typologies to more nuanced analyses of in-game decision making and behavioral clustering. 
Early frameworks, like \citet{bartle1996hearts}'s player types, provide an initial taxonomy of player motivations (achievers, explorers, socializers and killers in Multi-User Dungeons) and Bartle has since revisited his own framework acknowledging the constantly changing nature of these taxonomies as affordances of technology evolve~\cite{bartle2005virtual}.
Subsequent research has extended these ideas to players in modern team-based games, which consider both individual roles and group responsibilities ~\cite{eggert2015classification,jiang2021wide}.
Recent work has also examined the social dimensions of gameplay, demonstrating that even single-player games are often played collaboratively through practices like ``tandem play''~\cite{consalvo2017player}.

These taxonomies continue to be refined and expanded upon to fit different contexts. 
~\citet{jiang2021wide} examined how players prioritize choice breadth with conformity to community expectations, suggesting that a user's role selection can be balanced by both user preference and social pressures dictated by the meta-game. 
Studies of the gaming community itself also acknowledge that players may have different motivations that drive and shape their behavioral patterns, further corroborated by~\citet{bisberg2025taxonomy}.
~\citet{williams2008debunking} surveyed players in the massively multiplayer online game \textit{EverQuest 2} and found that most play for immersion and social interactions, but are primarily driven by in-game achievement. 

These insights highlight the diversity of play styles and lay the groundwork for our investigation into whether play styles shift or remain consistent when environmental or structural incentives change.

\subsection{Game Design and Behavioral Incentives}
\label{sec:game_design}
The design or structure of a game and the characteristics of its characters can influence player behavior. 
The \textit{Proteus effect} illustrates how player perception and ability to relate with avatars can profoundly affect their in-game decisions~\cite{li2012player,yee2007proteus,csengun2022players}. 

In competitive environments, players often have a shared goal of achieving performance-based success. 
Studies in gamification have shown that performance-based incentives (e.g., rewards and leader boards) can change an individual's behavior, although the presence and longevity of behavioral shifts are highly context-dependent and community-dependent~\cite{hamari2014gamification,deterding2011gamefulness}. 

The metagame, or \textit{meta} (the overarching strategies that are generally considered most effective), can also shape community perceptions of optimal play~\cite{debus2017metagames}. 
\citet{lee2017identifying} found that while there is a predominant meta strategy in League of Legends that is correlated with higher competitive success, there are select off-meta strategies that yield higher win rates. 
Game patches and balance updates can further influence the meta by introducing structural change aimed at maintaining game balance and user satisfaction~\cite{claypool2017patch,kica2016nerfs,claypool2015surrender}.
This can enable or restrict the range of viable roles, underscoring the tension between creative play and adherence to optimal strategies. 

Our work focuses on two games developed and maintained by \textit{Riot Games}: League of Legends and Teamfight Tactics. 
We examine the consistency of player flexibility across these games, which differ in genre and structure. 
The difference in structure creates different behavioral incentives for players, creating an ideal natural experiment for examining how players navigate different performance pressures. 

\subsubsection{League of Legends}
\label{sec:game_design_league}
League of Legends or \textit{League} is a Multiplayer Online Battle Arena game released by Riot Games in 2009~\footnote{\url{https://www.leagueoflegends.com/en-us/}}.
It is played in two teams of five, with each team typically consisting of five distinct roles: Top Lane, Middle Lane, Jungle, Bottom Lane (often referred to as an ADC, or Attack Damage Carry) and Support. 
Each role is generally associated with specific character -- or \textit{champion} -- types, with champion viability fluctuating based on game balance and the evolving meta.
Before a game begins, players select a primary and secondary role (with the option to ``fill'' any role); Riot then assigns roles to users and, while not strictly enforced, role assignments encourage a balanced team composition.  

Players select their champion during a pre-game draft phase, and in most modes, no champion can be selected more than once. 
The player then controls their chosen champion for that match from a top-down perspective. 
There are smaller objectives on the map to fight for that can temporarily or permanently increase a team's statistics. 
The primary objective of the game is to destroy the opposing team’s base, or \textit{Nexus}, while defending one's own. 
New champions are introduced on a semi-regular basis, and balance patches are scheduled every two weeks to adjust the game dynamics. 
Major structural changes are typically introduced at the beginning of a ``split'', with the number of splits increasing from a singular split to multiple splits a year over the last several years.  
As of this writing, Riot has released 170 playable champions available for users to purchase and play.

League's design requires players to commit to their champion selection before gameplay begins, with no opportunity to change during the match. 
This pre-game commitment structure, combined with the depth of mechanical mastery required for individual champions and the need for team coordination, creates incentives for players to specialize; players who develop expertise in a narrower set of champions are likely see increased performance than those who spread their practice across more champions~\cite{do2021machine,costa2021feature}.

\subsubsection{Teamfight Tactics}
\label{sec:game_design_tft}
Teamfight Tactics (TFT) was released by Riot Games in 2019 and is classified as an auto-battler or auto-chess game~\footnote{\url{https://teamfighttactics.leagueoflegends.com/en-us/}}. 
In TFT, eight players compete individually, each managing their own board during the planning phase by placing pieces or \textit{units} on their board -- many of these units are recognizable champions from League that may or may not have similar abilities or functions. 
Units vary in cost, with stronger units typically being more expensive. 
The economy system is contained within each game instance, meaning there is no way to convert real currency to game currency (gold); all players start with equivalent resources.

Players must both strategically spend gold to purchase units and buy experience points -- experience points will allow a user to level up and place additional units on the board. 
Players have a pool of gold they can use to purchase units from an individual shop displaying five random options, spend gold to ``reroll'' for new selections or level up for more team slots on their board. 
As a user levels up, the probability that more powerful units appear in a user’s shop increases.
One of the key features of TFT is that, for most units, all players draw from a shared pool with finite quantities. 
This means that as opponents purchase certain units, those units become increasingly scarce or unavailable to others. 

After each planning phase, players face off in one-on-one matches, where their units automatically execute actions based on preprogrammed behaviors. 
A player wins the round by eliminating all of their opponent’s units. 
Each player begins with 100 health and loses health points when they lose a round; elimination occurs when their health drops to zero, with the top four placements considered a ``win''. 
Unlike League, TFT operates in ``sets'' that are refreshed every few months, but TFT developers also release balance patches every two weeks to similarly adjust game mechanics. 
Each new set introduces new units, synergies, abilities and sometimes new mechanics that add complexity to the game while keeping the core gameplay intact. 

In contrast to League's pre-game commitment structure, TFT incorporates intentional randomness in unit availability, encouraging players to continuously adapt their strategies during gameplay. 
Because units are drawn from a randomized system through a shared pool, opponents' selections influence both the probability of finding the needed units and the optimal counter-strategies. 
This design creates incentives for flexibility: players who can pivot between compositions based on available units and opponent strategies tend to outperform those who rigidly commit to a single approach regardless of circumstance. 

These structural differences between League (rewarding specialization through pre-game commitment) and TFT (rewarding flexibility through dynamic adaptation) create an opportunity to examine how individual players navigate opposing performance incentives across different game environments.  
We approach this through Giddens' Structuration Theory, which provides a framework for understanding how individual agency and structural constraints mutually constitute behavior. 
To develop our hypotheses, we draw on research on the competence motivation to predict behavioral adaptation and personality stability to predict dispositional consistency across contexts.

\subsection{Self-Determination Theory and Player Motivation}
Self-Determination Theory (SDT) provides a framework that can be applied in games~\cite{tyack2020sdt} to understand why players might resist or embrace structural incentives; our first hypothesis draws on SDT's concept of competence.
SDT identifies three fundamental psychological needs: autonomy (having choice or volition), competence (exhibiting effectiveness or mastery) and relatedness (the feeling of connection to others)~\cite{ryan_self-determination_2000}.
While research has shown all three dimensions shape behavior in games~\cite{ryan_motivational_2006}, our study focuses specifically on competence as the motivational driver for performance-based success in competitive environments. 

The competence motivation manifests in competitive gaming as a driver for performance-based success, leading players to optimize their behavior for competitive achievement, or \textit{competitive success}~\cite{przybylski2010motivational}. 
In ranked competitive environments where performance is explicitly measured and rewarded through visible rank progression, the competence motivation should be particularly salient. 
Players seeking to demonstrate mastery and effectiveness, based on the competence motivation, should adapt their behavior to align with what the game's structural design rewards. 
We would expect players to specialize in contexts where specialization leads to success and diversify in contexts where flexibility leads to success.

Given the structural differences between League and TFT described in Section~\ref{sec:game_design}, competence-seeking players should adapt their behavior, with users exhibiting lower flexibility in League, where pre-game commitment and mastery depth reward specialization (see Section~\ref{sec:game_design_league}), and higher flexibility in TFT, where randomness and dynamic adaption reward flexibility (Section~\ref{sec:game_design_tft}). 

Because the competence motivation drives players to demonstrate effectiveness through performance-based success, we expect them to optimize their behavior to align with what each game's structure rewards. This leads to our first hypothesis:

\textbf{H1 (Competence-driven adaptation):} Flexibility will be negatively associated with competitive success in League and positively associated with competitive success in TFT, reflecting each game's structural incentives.

\subsection{Personality Psychology and Cross-Context Behavioral Stability}
While the competence motivation predicts behavioral adaptation to optimize performance, research in personality psychology demonstrates that core dispositional preferences persist across contexts as stable individual differences. 
This addresses the classic person-situation debate in personality research: specific behaviors vary substantially across situations, but stable individual differences persist in how people characteristically respond to different contexts~\cite{mischel1995cognitive,fleeson2001toward}.
Modern research resolves this paradox by showing that behavioral signatures, or distinctive patterns of behavioral variation across situations, remain stable even when surface behaviors seem to change.

Prior meta-analyses show that personality trait consistency is substantial and increases with age~\cite{roberts2000rank,caspi2005personality}. 
This stability extends to behavioral tendencies;~\citet{epstein1979stability} found that when behavior is aggregated across multiple instances, cross-situational consistency increases dramatically. 
This \textit{aggregation principle} is directly relevant to our study design, where we track players who have played a minimum number games per title to reveal stable dispositional patterns.

~\citet{fleeson2001toward}'s density distributions approach demonstrates that while an individual's transient behavior may vary, individual differences in core behavioral tendencies remain highly stable across different contexts. 
Even in games such as League and TFT, although a player might select different champions or compositions from match to match, respectively, their average proclivity toward breadth versus focus should persist across different gaming environments. 
Mischel and Shoda's CAPS theory~\cite{mischel1995cognitive} formalizes this through ``behavioral signatures,'' where if-then patterns reflect underlying dispositional tendencies that remain stable.
In gaming contexts specifically, research demonstrates that real-world personality traits predict virtual world behavior~\cite{yee2011introverted}, confirming that dispositional preferences extend to digital environments.

This research predicts that relative positions on flexibility versus specialization should persist across games, even if absolute flexibility levels shift somewhat in response to structural incentives (H1). 
Players who are more flexible than average in one game should tend to be more flexible than average in the other because their underlying behavioral signature -- their characteristic way of approaching competitive challenges -- remains stable. This leads to our second hypothesis:

\textbf{H2 (Dispositional persistence):} Individual players will maintain consistent relative flexibility patterns across games, such that flexibility in one game predicts flexibility in the other independent of competitive success, reflecting stable personality-level preferences.

\subsection{Agency versus Structure}
\label{sec:background_agency_vs_structure}
H1 and H2 generate competing predictions: the competence motivation predicts behavioral adaptation to structural incentives, while personality research predicts dispositional consistency across contexts. 
Giddens' Structuration Theory~\cite{Giddens1979agency} provides a framework for understanding how both patterns can occur simultaneously, with the balance depending on individual motivation.

A core debate in sociology is whether \textit{agency} -- an individual's preference or capacity to make decisions -- or \textit{structure} -- the external system and its rules -- plays a more decisive role in determining behavior. 
~\citet{Giddens1979agency}'s work on Structuration Theory introduces the concept of the duality of structure, where he posits that while structure may shape behaviors, an individual's agency and choice can also reshape those very structures. 
This perspective is supported by reviews and case studies of this interplay between agency and structure, reaffirming mutual inclusivity~\cite{tan2011understanding, karp1986agency,stones2017structuration}. 
Giddens also argues that individuals are knowledgeable agents who reflexively monitor their actions and develop routines through repetition~\cite{Giddens1979agency}.  
These routines operate through \textit{practical consciousness}, or tacit knowledge that agents use but cannot easily describe, and persist across contexts because they are deeply ingrained through prior experience. 

Research on habit formation echoes Giddens' Structuration Theory. 
~\citet{lally2013promoting} and ~\citet{ersche2017habit} argue that habits are automatic behavioral responses to situational stimuli through repetition, where conscious action transforms into subconscious routine. 
Changing behavior requires deliberate \textit{implementation intention}, indicating that while structure plays a crucial role in habit formation, individual intention or agency is also key to triggering behavioral change~\cite{lally2013promoting}. 

However, Giddens also emphasizes the concept of \textit{reflexive monitoring}, defined as ongoing self-awareness of one's own actions and the ability to adjust one's behavior as needed in specific circumstances. 
While routines and habits (like flexibility preferences) may persist, individuals with sufficient motivation can purposefully evaluate their situation and make changes to their behavior even when it conflicts with established patterns~\cite{Giddens1979agency,giddens1984constitution}. 
This implies that while prior dispositions exist, conscious adaptation can precipitate behavioral change. 
The duality of structure means that both adaptation to structure (supporting H1) and persistence through agency (supporting H2) can occur simultaneously, although the balance between the influence of the two will shift based on individual motivation and the salience of environmental incentives -- or in our case, performance goals.

In the context of our study, Structuration Theory suggests that flexibility patterns will operate as routinized practices for most players, persisting across games as practical consciousness and supporting H2.
However, elite players who are more motivated by competitive success will engage in more reflexive monitoring, resulting in a conscious evaluation and realignment of their levels of flexibility with a game's structural incentives. 
The completeness of this adaptation will be tempered by the duality of structure, as dispositional preferences will still play a role in constraining the extent to which this behavior can change, even under conscious deliberation. 

This leads to our third hypothesis, which reconciles H1 and H2: 

\textbf{H3 (Partial adaptation by motivated players):} Elite players, for whom competitive success is more salient, will show greater adaptation to structural incentives (H1) compared to non-elite players. 
This adaptation, however, will not be a complete realignment, with baseline dispositional tendencies (H2) still evident across both groups. 
Reflexive monitoring enables conscious evaluation and partial modification of practices when motivation is high, but the duality of structure constrains complete behavioral change.

\section{Methods}
Because League of Legends and Teamfight Tactics share a common desktop client, players are able to download, update and access both games with minimal additional effort and through the same account information. 
The shared ecosystem allows us to track the same user across both games through a common username. 
This reduces self-selection bias that often arises when users must opt in to disclose information about multiple platforms, as those who do choose to do so may not be representative of the general population~\cite{Heckman2018bias,bethlehem2010bias}. 
While this design choice limits our analysis to games within the Riot ecosystem, it provides a methodologically rigorous approach to studying cross-game behavioral consistency by ensuring that observed patterns reflect genuine individual stability rather than differences between user populations who select into different games. 
Individual-level behavioral data tracking the same users across multiple games without self-selection bias is rare in games research due to technical and access limitations, making this dataset a valuable opportunity to empirically test agency-structure dynamics that are typically confounded by selection effects.

\subsection{User Sampling}

\subsubsection{Balancing Based on Skill and Rank}
Our study examines the consistency -- or lack thereof -- in player behavior, focusing on the breadth of a user's play style and their willingness to adapt strategies in competitive environments. 
Competition is inherently motivated by performance-based success, which may differentiate elite players from non-elite ones and lead to distinct behavioral patterns between the two groups. 

We categorize players based on Riot Games' designation of the top three tiers (Master, Grandmaster and Challenger) as \textit{Apex tiers}. 
Players in these tiers represent less than 1\% of the total user base and we label these users as \textit{elite}, while players outside of these tiers are considered \textit{non-elite}.
This binary classification serves several purposes: Apex tier players face distinct competitive structures (including, during this dataset's timeframe, rank decay penalties for inactivity and stricter matchmaking), are more likely to be professional or semi-professional players for whom competitive success is paramount, and play significantly more games on average (see Table~\ref{tab:league_tft_seeds_comparison}). 
From an SDT perspective, elite players also represent individuals with high competence, allowing us to test whether competence satisfaction moderates autonomy-driven preferences.
Practical constraints with API rate limitations also shaped our sampling strategy. 
We intentionally over-sampled elite players over non-elite players to ensure that our final dataset collection is balanced for robust analysis in a performance-based ecosystem. 

We use a player's win rate as a proxy for \textbf{competitive success}. 

\subsubsection{Identifying and Tracking Users}
We leveraged Riot Games' public API~\footnote{\url{https://developer.riotgames.com/apis}} along with several popular third-party statistic tracking websites -- OP.GG\footnote{\url{https://op.gg}} for League data and LoLchess\footnote{\url{https://lolchess.gg}} for TFT -- to compile our list of users to track to construct our longitudinal dataset. 
While OP.GG provides a comprehensive list of ranked players, LoLchess only displays the top $1000$ users per rank division (e.g., Platinum 1, Platinum 2). 
Although only non-elite or non-apex ranks have rank divisions (each non-apex rank has four subdivisions), we are still able to retrieve a representative sample of users across all subdivisions in a rank.

We retrieved the available North American leader boards for League and TFT from their respective websites on May 15, 2023. 
A player's \textit{seed game} was determined by the leader board from which they were extracted and their \textit{skill level} was classified as elite or non-elite based on their competitive rank at the time of collection. 
The ranking system, from lowest to highest, is as follows: Iron, Bronze, Silver, Gold, Platinum, Diamond, Master, Grandmaster and Challenger. 
(Note: a new rank, Emerald, was introduced later between Diamond and Platinum; however, we do not include this rank in our discussions since it was rank was not yet implemented during our initial dataset.)
Due to rate limitations imposed by Riot's API, we sampled and tracked a portion of the leader board, using a higher sampling percentage for elite players to maintain a balanced dataset between the number of elite players and non-elite players in our final dataset (see Table~\ref{tab:user_distribution} for distributions). 
It is possible for a user to be present in our list of TFT seed users and League seed users.
In these cases, we count this user twice according to their measured ranking in each game. 

Table~\ref{tab:user_distribution} summarizes the user percentages and distributions for the full dataset and our final dataset that includes only users who have played a minimum of 50 Solo Queue ranked games in each title. 
We track the same users in both titles, and our final dataset only has one user who is thus double counted after being present in the ranked ladders of both games (non-Elite in both League and TFT). 

\begin{table*}[ht]
\centering
\begin{tabular}{c|c|c|c|c|c}
\textbf{Game} & \textbf{Elite/Non-Elite} & \textbf{Rank} & \textbf{Sample \%} & \textbf{Total Users} & \textbf{Num Users (50+ games)} \\ \hline
\multirow{3}{*}{League} & \multirow{2}{*}{Elite} & Challenger, Grandmaster & All & \multirow{2}{*}{4018}  & \multirow{2}{*}{1996} \\ \cline{3-4}  
 &  & Master & 50\% &  &  \\ \cline{2-6}
 & Non-Elite & Diamond - Iron & 0.5\% & 5621 & 1575 \\ \hline
\multirow{3}{*}{TFT} & \multirow{2}{*}{Elite} & Challenger, Grandmaster & All & \multirow{2}{*}{1729} & \multirow{2}{*}{563} \\ \cline{3-4}
 &  & Master & All &  &  \\ \cline{2-6}
 & Non-Elite & Diamond - Iron & 50\% & 2854 & 696 \\ 
\end{tabular}
\caption{User skill distribution and user counts for the full dataset and final dataset (users who have played 50+ ranked games in each title).}
\label{tab:user_distribution}
\end{table*}

\subsection{Data Collection}
\subsubsection{Quasi-Real-Time Data Infrastructure}

Each user is uniquely identified by a Player Universally Unique Identifier (PUUID) within the Riot ecosystem and API, which allowed us to continuously track users regardless of any username changes. 
The Riot API provides data on a user's most recent 100 League games and 200 TFT games. 

We developed a quasi-real-time tracking infrastructure to capture player activity, cycling through the list of users and querying Riot's servers for each user's most recent games. 
If a new match is detected, we used Riot's API to request that match's data; this enabled us to construct a fairly comprehensive play history for each user throughout our collection period. 

\subsubsection{Competitive Queues and Game Modes}
While competitive success is not the central focus of this paper, it still plays an important role in our models. 
Both League and TFT offer multiple game modes with varying levels of competitiveness. 
In League, ranked modes include Solo Queue and Flex Queue -- Solo Queue is the most competitive and it limits players to queuing with one teammate of a similar rank (and restricts team queuing altogether for higher ranks), while Flex Queue is perceived as less competitive as it has significantly fewer restrictions. 
In TFT, despite being an individual game, players can queue with up to two other players; higher-ranked players are also restricted from queuing with others. 
Our analyses focuses on games played in the most competitive ranked queue -- Solo Queue -- for both titles. 

\subsubsection{Dataset Composition}
Our dataset actively tracked users from May 22, 2023 to July 22, 2024, and the earliest TFT game took place in September 2019 and League game in December 2020. 
We also only included users who have played at least 50 Solo Queue games in both League and TFT, totaling a minimum of 100 games per user (see Table~\ref{tab:user_distribution}). 
This resulted in a final dataset that contains 4,830 users.
The game threshold allowed us to capture well-developed play styles and eliminate users with insufficient exposure or investment in either game. 
This time frame spanned several TFT sets and League splits; while rankings are reset during a new set or season, a user's skill classification was based off of their rank when the collection began. 

\subsection{Defining User Flexibility}

Flexibility in gameplay is the sum of many factors, encompassing not only the diversity of decisions a player makes but also the ability to adapt those decisions over time~\cite{debus2017metagames}. 
The game design itself can also influence how flexibility is expressed. 
In League, players commit to champion selections during a pre-game draft, which constrains their in-game adaptability. 
TFT, on the other hand, does not typically require pre-game commitments and generally allows players greater freedom to adjust strategies during the game (although players can enter the game with preconceived strategies). 

Riot's API only returned the end state of each League or TFT game; therefore we quantified flexibility using these end-game statistics that reflect a user's final play style choices.
For simplicity, we leveraged predefined game design decisions that categorize user choice into distinct roles or compositions. 

\subsubsection{Flexibility in League of Legends}
We measured a player's flexibility in League based on the breadth and frequency of \textit{functional play styles} a user chose to play over all games in our data collection. 

Riot Games has designed champions that fit into six primary \textit{play style} roles: Assassins, Fighters, Mages, Marksmen, Supports and Tanks. 
Each champion is assigned a primary role, and in some cases, will have a secondary role; this results in 21 order-agnostic combinations of play styles. 
We define a champion's \textit{functional play style} as this order-agnostic combination of play style roles, which defines how a champion is intended to be played within the game. 

We quantify a user's play style flexibility from the distribution of functional play styles that a user plays, loosely inspired by entropy calculations. 
A higher flexibility score corresponds to greater user flexibility while a lower score indicates that a user is more narrow in play style. 
This method allowed us to capture both role diversity and consistency of an individual's play style tendencies, and we describe the calculations in further detail below. 

\textbf{Role Proportions:} 
We first calculated (see equation~\ref{eq:pi}) the proportion of games that a user plays using each functional play style $i$, where $N_i$ represents the number of games a user plays a specific functional play style and $N_{games}$ is the total number of games played. 

\begin{equation}
    P_{i} = \frac{N_i}{N_{games}}
    \label{eq:pi}
\end{equation}

\textbf{Preferred Play Style:} 
We then identified the most frequently played play style, $P_{max}$ through the highest $P_i$ value (see equation~\ref{eq:pmax}).

\begin{equation}
    P_{max} = max(P_1,P_2...P_{21})
    \label{eq:pmax}
\end{equation}

\textbf{Flexibility Score:}
We captured the overall balance in play style engagement by computing the normalized proportion deviation $P_{norm}$ as the average absolute difference between the maximum proportion $P_{max}$ and each play style proportion (see equation~\ref{eq:flex}). 

A lower $P_{norm}$ suggests a more uniform distribution across play styles, which indicates higher flexibility. 

Finally, we defined the flexibility score as $F_{league}$ for League as $1-P_{norm}$ such that a higher $F_{league}$ indicates greater flexibility, meaning the player uses a wider breadth of roles (equation~\ref{eq:flex}). 

\begin{equation}
    F_{league} = 1 - P_{norm} \mbox{\quad with \quad } P_{norm}=\frac{1}{21}\sum^{21}_{i=1}|P_{max}-P_i|
    \label{eq:flex}
\end{equation}


\subsubsection{Flexibility in Teamfight Tactics}
TFT's game design and structure requires a slightly modified approach to quantify flexibility, as player decisions revolve around creating different \textit{compositions} of units that are then defined by \textit{origins} and \textit{classes}. 

Each unit is defined by a combination of origins and classes. 
Origins typically provide simpler buffs to unit statistics at certain breakpoints (e.g., increased health or armor when a certain number of unique units of the same origin are played) while classes offer more complex and synergistic effects at their breakpoints.  
We categorized a player's \textit{board composition} based on the \textit{origin} and \textit{class} that reached the highest percentage of possible breakpoints, with any ties broken randomly. 
Because each set has different units, origins and classes, we averaged a user's per-set flexibility scores resulting in an overall flexibility score for each user. 
Our method for calculating flexibility in TFT is similar to that used for League, with slight adjustments to account for TFT's mechanics and is detailed below. 

\textbf{Total Composition Categories:}
We first calculated the total number of possible pairs of origins and classes for each set, where $N_{origins}$ is the number of distinct origins and $N_{classes}$ the number of distinct classes. 
Equation~\ref{eq:ncomp} captures the pairwise combinations of individual origins and classes, adding one to account for the lack of origins or classes. 

\begin{equation}
N_{comps} = N_{origins} + N_{classes} + (N_{origins} \times N_{classes}) + 1
\label{eq:ncomp}
\end{equation}

\textbf{Composition Proportions:}
For each composition $i$, let $N_i$ be the number of games in which composition $i$ was played and $N_{games}$ be the total number of TFT games played. 
The proportion is given in equation~\ref{eq:pi_tft}. 

\begin{equation}
P_i = \frac{N_i}{N_{games}}
\label{eq:pi_tft}
\end{equation}

\textbf{Preferred Composition:}
We then determined the user's most frequently played composition, $P_{max}$ through the highest $P_i$ value (see equation~\ref{eq:pmax_tft}).

\begin{equation}
P_{max} = \max(P_1, P_2, \dots, P_{comps})
\label{eq:pmax_tft}
\end{equation}

\textbf{Flexibility Score}
Just as with League, we calculated the average deviation from the preferred composition and took its complement (equation~\ref{eq:flex_tft}). 
A higher $F_{tft}$ indicates that the player plays a more even distribution of compositions, reflecting higher flexibility. 

\begin{equation}
F_{tft} = 1 - \frac{1}{N_{comps}} \sum_{i=1}^{N_{comps}} \left| P_i - P_{max} \right|
\label{eq:flex_tft}
\end{equation}


\subsection{Model Features}
Our analysis focuses specifically on Solo Queue games (the most competitive mode in both titles), ensuring that success is consistently defined in both games. 
Each user's \textit{seed skill} is categorized as \textit{elite} (top three ranked tiers comprising of less than 1\% of the user base) or \textit{non-elite} based on their recorded rank at the beginning of data collection.
We briefly describe the features below and a summary of the features can be found in Table~\ref{tab:model_features}; a detailed description can be found in Appendix~\ref{sec:feature_descriptions}.

\subsubsection{Performance-Based Features}
We defined \textit{competitive success} as a player's win rate in Solo Queue in a game and a user's most played functional play style or composition as their preferred play style or composition.
We also considered the difference between a user's win rate when playing their most common functional role (or composition) and their overall win rate. 

\subsubsection{User Flexibility Features}
A user's flexibility score was computed separately for both League and TFT. 
TFT borrows many of its assets from League; because of this, we also included whether a user's preferred play style (based on the League definition of play style) was aligned across the two games. 
We further accounted for intra-role flexibility by including the number of champions or units available in a user's preferred play style or composition. 

\subsubsection{Engagement Features}
Comparable general engagement metrics in both games include the fraction of Solo Queue games played (out of all games played), the percentage of days on which only one game is played versus both games and metrics related to session length. 
Prior work indicates that performance may decline after a certain threshold of consecutive games in a session~\cite{sapienza2018individual}. 
Thus, we also measured the average number of consecutive games played and identified in which game a user has a higher average daily streak of consecutive games.  
Finally, we captured overall daily engagement through metrics such as the average number of games played per day, average games per active day and trends in daily play rates over time. 

\begin{table*}[ht]
\centering
\begin{tabular}{c|c|c|c}
\textbf{Feature} & \textbf{League} & \textbf{TFT} & \textbf{General} \\ \hline
Win Rate & \checkmark & \checkmark &  \\
Avg Daily Session & \checkmark & \checkmark &  \\
Avg Daily Games & \checkmark & \checkmark &  \\
Flexibility Score & \checkmark & \checkmark &  \\
Preference Win Rate Difference & \checkmark & \checkmark &  \\
Number of Options in Preferred Play Style/Composition & \checkmark & \checkmark &  \\
Temporal User Engagement & \checkmark & \checkmark &  \\
\% all games played in Solo Queue & \checkmark & \checkmark &  \\
Avg Daily Streak Length** & \checkmark & \checkmark &  \\
\% days playing only one game** & \checkmark & \checkmark &  \\

\% days both games played &  &  & \checkmark \\
League Streak Greater &  &  & \checkmark \\
Role Pref Same &  &  & \checkmark \\
Seed Skill (Non-Elite)*  &  &  & \checkmark \\
Seed Game (TFT)*  &  &  & \checkmark \\

\end{tabular}
\caption{Summary of model features. * indicates dummy variable for categorical variables. ** indicates features that were not included in the MVLR and LMEM models due to high collinearity.}
\label{tab:model_features}
\end{table*}

\section{Results}
\subsection{General Dataset Statistics}
Our final dataset consisted of users who have played at least 50 Solo Queue games in each game (League and TFT), resulting in a minimum of 100 total competitive games. 
We selected this threshold because it represents half of the most recent League games that we are able to retrieve for each user, and ensures that each user has had time to learn the game and develop a distinguishable win rate and play style. 
Table~\ref{tab:alldata_metrics} presents overall metrics of the dataset while Table~\ref{tab:league_tft_seeds_comparison} disaggregates these statistics by a user's seed game and seed skill.

\begin{table}[ht]
\centering
\begin{tabular}{c|c}
\textbf{Game Metric} & \textbf{Value}\\ \hline
Solo Queue Win Rate (League) & 50.55\% \\ 
Solo Queue Win Rate (TFT) &  51.80\% \\ 
Avg. Daily Games Played (League)& 0.81 \\ 
Avg. Daily Games Played (TFT)& 0.45 \\ 
Games per Daily Session (League) & 3.55 \\ 
Games per Daily Session (TFT) & 3.09 \\ 
\% Days Played Only League & 49.30\% \\ 
\% Days Played Only TFT & 33.12\% \\ 
\% Days Played Both Games & 12.11\% \\ 
\end{tabular}
\caption{Comparison of game-specific metrics (50\%-tile) for League and TFT.}
\label{tab:alldata_metrics}
\end{table}

\begin{table*}[ht]
\centering
\begin{tabular}{c|cc|cc}
\multirow{2}{*}{\textbf{Metric}} & \multicolumn{2}{c|}{\textbf{League}} & \multicolumn{2}{c}{\textbf{TFT}} \\ 
 & \textbf{Elite} & \textbf{Non-Elite} & \textbf{Elite} & \textbf{Non-Elite} \\ \hline
Solo Queue Win Rate (League) & 51.08\% & 49.98\% & 50.85\% & 50.15\% \\  
Number of Daily Games (League) & 1.39 & 0.59 & 0.49 & 0.46 \\ 
Number of Games Played in a Daily Session* (League) & 4.23 & 3.12 & 3.47 & 3.04 \\ 
Solo Queue Win Rate (TFT) & 52.18\% & 49.66\% & 53.97\% & 51.95\% \\  
Number of Daily Games (TFT) & 0.37 & 0.30 & 2.58 & 0.62 \\ 
Number of Games Played in a Daily Session* (TFT) & 3.10 & 2.60 & 5.83 & 2.93 \\
\% Days Played Only League & 59.24\% & 50.65\% & 27.56\% & 33.85\% \\
\% Days Played Only TFT & 26.28\% & 35.75\% & 45.62\% & 42.59\% \\ 
\% Days Played Both Games & 10.05\% & 9.77\% & 21.27\% & 15.28
\end{tabular}
\caption{Comparison of game-metrics disaggregated by game and skill seed. Metrics reported are the 50 \%-tile values.}
\label{tab:league_tft_seeds_comparison}
\end{table*}

In looking at the daily distribution of games, elite players are more likely to play both games on the same day compared to non-elite players, particularly among TFT seed players. 
This pattern likely reflects the historical context in which many players initially played League (which was released nearly 10 years prior to TFT) before also adopting TFT. 

We also calculated several other metrics -- total daily games, average games per active day and games per daily session -- to capture both overall engagement and session length. 
Most players play about three games per daily session, consistent with prior findings~\cite{sapienza2018individual}; however, elite TFT players engage with nearly twice as many games per daily session, which may be due to TFT's design and individual nature that allows players to exit a game immediately after a loss without affecting anyone else in the game -- unlike in League (team-based), where leaving early incurs penalties of increasing severity. 
Finally, average win rates in both games hover near 50\%. 
This is a direct consequence of skill-based matchmaking, whose explicit goal is to assemble lobbies of comparable ability, where each League team and each individual in TFT has a roughly even chance of victory at the beginning of a match.

\subsection{Modeling Competitive Success}
\textit{Competitive success} -- which we defined as a player's win rate in Solo Queue games -- is a critical performance-based metric that influences player behavior, especially in competitive environments. 
H1 predicted that League and TFT's distinct structural incentives influence the relationship between player flexibility and competitive success, with flexibility negatively associated with competitive success in League and positively associated in TFT.
To test this hypothesis, we explored how engagement and flexibility features correlate with competitive success and train multivariate linear regression (MVLR) models for both League and TFT. 

As expected and as corroborated by prior work, we found that in-game performance in our initial models had high predictive power~\cite{silva2018continuous,garcia2023machine}. 
We ultimately excluded these variables to focus on the effects of flexibility and engagement on competitive success. 
This resulted in a decrease in $R^2$ values, but the models still allowed us to identify key correlations between our features and competitive success. 
Table~\ref{tab:mvlr} summarizes the results from our models predicting competitive success in League and TFT. 

The results showed that elite players generally achieve higher competitive success, and strongly supports H1. 
In League, increased flexibility, or a more uniform distribution of play styles, is negatively correlated with competitive success -- this suggests that specialization is beneficial in performance-based success. 
The opposite is true for TFT, where increased flexibility is positively correlated with competitive success, indicating that the use of a broader range of strategies is advantageous. 
This operationalizes the variation of structure, while the players and their internal preferences remain consistent. 

\begin{table*}[ht]
\centering
\begin{tabular}{c|ccc|ccc}
\multirow{2}{*}{\textbf{Feature}} & \multicolumn{3}{c|}{\textbf{Model 1 (League Success)}} & \multicolumn{3}{c}{\textbf{Model 2 (TFT Success)}} \\ 
 & \textbf{Coefficient} & \textbf{Std Err} & \textbf{P Value} & \textbf{Coefficient} & \textbf{Std Err} & \textbf{P Value} \\ \hline
League Flexibility & -0.009 & 0.003 & 0.001\textsuperscript{***}& 0.003 & 0.004 & 0.375 \\
TFT Flexibility & -0.011 & 0.004 & 0.012\textsuperscript{*} & 0.035 & 0.006 & 0.000\textsuperscript{***} \\
Seed Skill (Non-Elite) & -0.008 & 0.001 & 0.000\textsuperscript{***} & -0.030 & 0.002 & 0.000\textsuperscript{***} \\
Seed Game (TFT) & 0.001 & 0.001 & 0.588 & 0.016 & 0.002 & 0.000\textsuperscript{***} \\
\end{tabular}
\caption{Select MVLR Coefficients, Standard Errors, and P Values for Model 1 (League Competitive Success) and Model 2 (TFT Competitive Success). Note: *$p \leq 0.05$,  ***$p \leq 0.001$}
\label{tab:mvlr}
\end{table*}

\subsection{Cross Game Flexibility}
\begin{figure*}[ht]
    \centering
    
    \begin{subfigure}{0.45\textwidth}
        \centering
        \includegraphics[trim=0 0 0 35, clip,width=\linewidth]{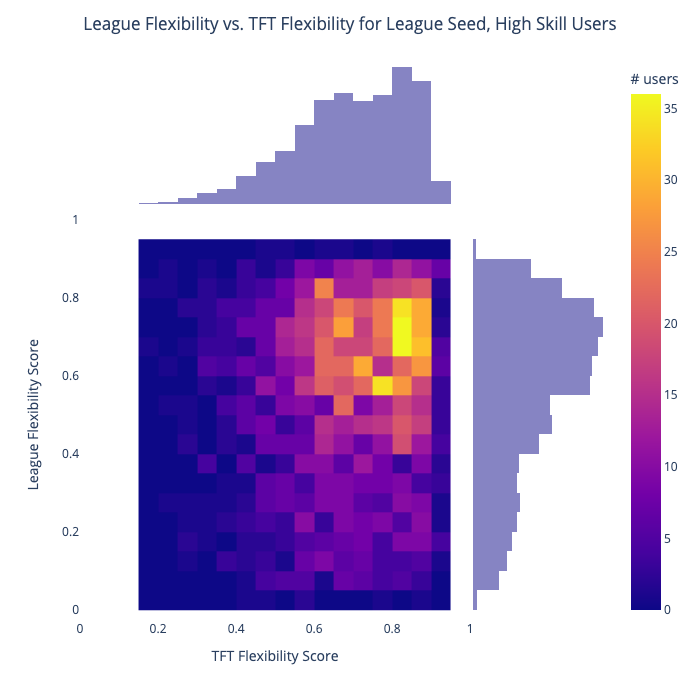}
        \caption{League seeds, Elite players}
        \label{fig:league_seeds_elite_map}
    \end{subfigure}
    \hfill
    \begin{subfigure}{0.45\textwidth}
        \centering
        \includegraphics[trim=0 0 0 35, clip,width=\linewidth]{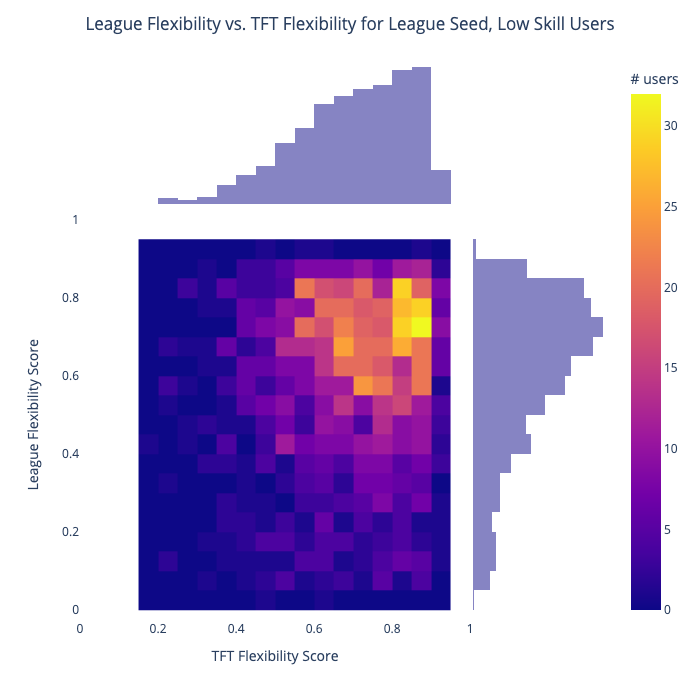}
        \caption{League seeds, Non-Elite players}
        \label{fig:league_seeds_nonelite_map}
    \end{subfigure}
    
    \vspace{0.5cm}
    
    \begin{subfigure}{0.45\textwidth}
        \centering
        \includegraphics[trim=0 0 0 35, clip,width=\linewidth]{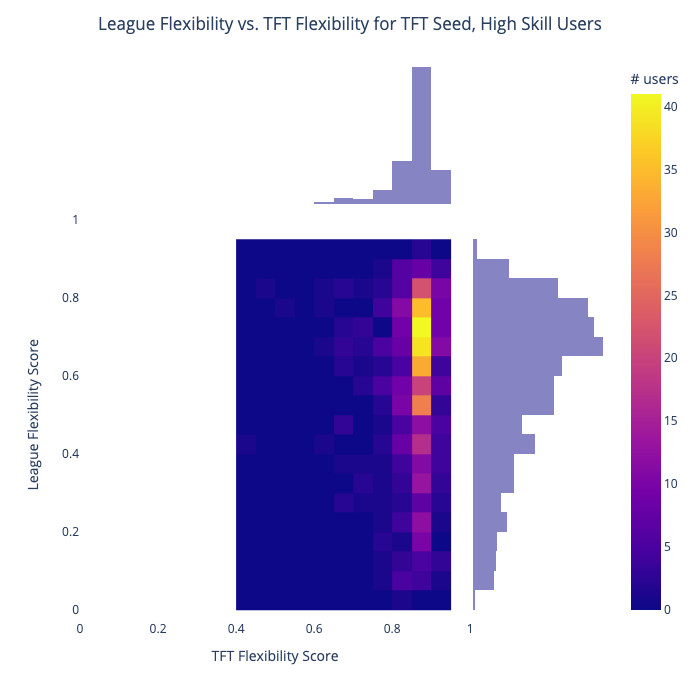}
        \caption{TFT seeds, Elite players}
        \label{fig:tft_seeds_elite_map}
    \end{subfigure}
    \hfill
    \begin{subfigure}{0.45\textwidth}
        \centering
        \includegraphics[trim=0 0 0 35, clip,width=\linewidth]{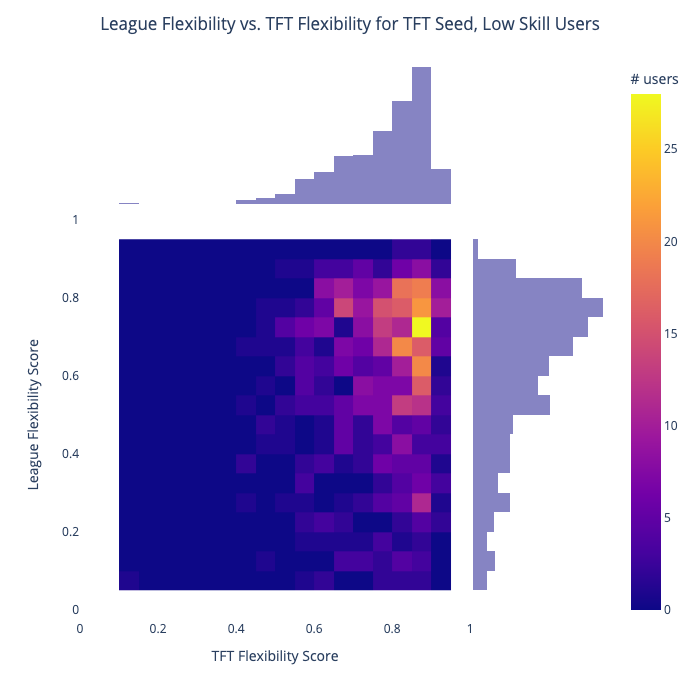}
        \caption{TFT seeds, Non-Elite players}
        \label{fig:tft_seeds_nonelite_map}
    \end{subfigure}
    
    \caption{A 2x2 grid of heatmaps comparing TFT flexibility (x-axis) and League flexibility scores (y-axis), disaggregated by seed game by row (League seeds - top, TFT seeds - bottom) and seed skill by column(elite - left, non-elite - right). This enables us to compare the relative flexibility of the categories of users to each other. }
    \label{fig:flexibility_map}
\end{figure*}

While H1 demonstrated that TFT and League reward opposing behavior, H2 predicts that individual dispositional preferences will persist across game contexts. 
We examine whether players maintain consistent relative flexibility patterns across League and TFT despite these opposing performance incentives. 
We calculated users' flexibility scores for TFT and League and generated heat maps to visualize the relationship between TFT (x-axis) and League (y-axis) flexibility scores disaggregated by a user's seed game and seed skill. 
This results in a 2 by 2 matrix found in Figure~\ref{fig:flexibility_map}. 

The heat maps show that elite players tend to optimize their behavior for their seed game compared to their non-elite counterparts: elite League seeds are more specialized in League (Figure~\ref{fig:league_seeds_elite_map}) while elite TFT seeds exhibit higher flexibility (Figure~\ref{fig:tft_seeds_elite_map}) in TFT.  

While we found that elite players are generally more competitively successful in both games regardless of seed game, elite players do not always adapt their play styles when switching to their non-seed game. 
For example, elite League seeds (Figure~\ref{fig:league_seeds_elite_map}) tend to be slightly less flexible in TFT compared to non-elite League seeds (Figure~\ref{fig:league_seeds_nonelite_map}).
This provides initial support for H2, indicating that while competitive success can drive behavior, behavioral trends in one game persist and are evident in cross-game behavioral patterns.

\subsection{Predicting User Flexibility}
Building on our finding that game structure influences the flexibility-success relationship differently in each game (H1), we now test H2 and H3 more rigorously by examining what factors predict player flexibility.
H2 predicts that cross-game flexibility will be a strong predictor of individual behavior independent of competitive success; H3 predicts that elite players will show different patterns than non-elite players, reflecting partial adaptation. 
Instead of using competitive success as the dependent variable, we now predict a user's flexibility score in each game.

\subsubsection{League of Legends Flexibility}
\label{sec:pred_league_flex} 
We modeled League flexibility using MVLR as a baseline, linear mixed-effects models (LMEM), gradient boosting machines (GBM), kernel regression and neural networks. 
Although we expected LMEMs to improve baseline performance by accounting for systematic differences between skill levels (elite vs. non-elite) and seed game (League vs. TFT seeds) as fixed effects, LMEMs performed worse than our baseline; non-linear and non-parametric models like GBM, kernel regression and neural networks resulted in better model performance.  
In particular, kernel regression with a Laplacian kernel achieved the best performance with an $R^2$ value of 0.2235 (Table~\ref{tab:league_flex_models}).
Hyperparameters were tuned using grid search and the Optuna framework~\cite{akiba2018optuna} and model performance was evaluated using 10-fold cross-validation. 

\begin{table*}[ht]
\centering
\begin{tabular}{l|c|c|c|c}
\textbf{Model Type} & \textbf{Fixed Effect} & \textbf{Kernel Type} & \textbf{MSE} & \textbf{$R^2$} \\ \hline
MVLR & -- & -- & 0.042 & 0.0215 \\
LMEM & Seed Skill  & -- & 0.0415 & 0.0069 \\
LMEM & Seed Game & -- & 0.0412 & 0.0135 \\
LMEM & Seed Skill  + Seed Game & -- & 0.0416 & 0.0135 \\ 
GBM & -- & -- & 0.0341 & 0.2093 \\
\textbf{Kernel Regression }& \textbf{--} & \textbf{Laplacian} & \textbf{0.03501} & \textbf{0.2235} \\
Neural Networks & -- & -- & 0.0354 & 0.1526 \\
\end{tabular}
\caption{Comparison of model results for predicting League flexibility. Kernel regression with a Laplacian kernel is the best-performing model. }
\label{tab:league_flex_models}
\end{table*}

\subsubsection{Teamfight Tactics Flexibility}
We applied a similar modeling approach to predict TFT flexibility, with League flexibility included as a feature when appropriate. 
Kernel regression with a Laplacian kernel again emerged as the best-performing model ($R^2$ = 0.4551, see Table~\ref{tab:tft_flex_models}). 

\begin{table*}[ht]
\centering
\begin{tabular}{l|c|c|c|c}
\textbf{Model Type} & \textbf{Fixed Effect} & \textbf{Kernel Type} & \textbf{MSE} & \textbf{$R^2$} \\ \hline
MVLR & -- & -- & 0.0131 & 0.3760 \\
LMEM & Seed Skill  & -- & 0.0146 & 0.3093 \\
LMEM & Seed Game & -- & 0.0151 & 0.2837 \\
LMEM & Seed Skill  + Seed Game & -- & 0.0151 & 0.2838 \\
GBM & -- & -- & 0.0121 & 0.4440 \\
\textbf{Kernel Regression} & \textbf{--} & \textbf{Laplacian} & \textbf{0.0116} & \textbf{0.4551} \\
Neural Networks & -- & -- & 0.0119 & 0.4375 \\
\end{tabular}
\caption{Comparison of model results for predicting TFT flexibility. Kernel regression with a Laplacian kernel is the best-performing model.}
\label{tab:tft_flex_models}
\end{table*}

\subsubsection{Comparing Model Performance}

While our baseline MVLR models performed poorly relative to the non-linear models, TFT flexibility was better predicted by engagement features compared to League flexibility. 
This discrepancy may be a result of the extent of game decisions that were captured in our data -- League involves complex temporal decision-making that may be less evident from end-game snapshots, while TFT's design allows these snapshots to better capture in-game decision dynamics. 

For both titles, kernel regression with a Laplacian kernel provided the best predictive performance. 
We used SHAP visualizations to interpret the models to determine the importance of individual features for model performance~\cite{lundberg2017shap}), and we describe our findings in more detail in the following section. 
All hyperparameters can be found in Appendix~\ref{sec:appendix_hyperparams}.

\begin{figure*}[!tb]
    \centering
    \includegraphics[width=\textwidth,keepaspectratio]{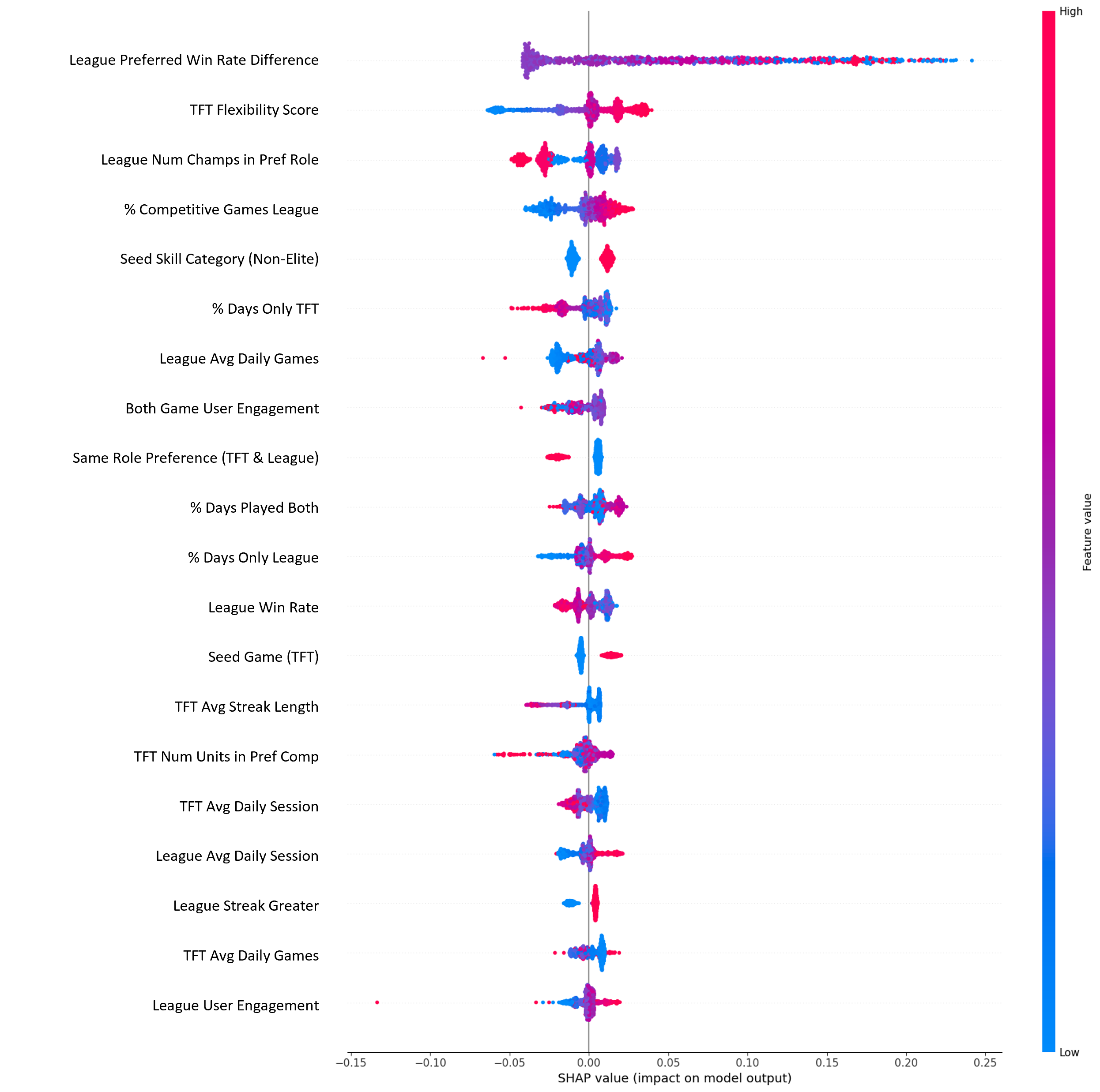} 
    \caption{SHAP visualization with top 20 features for our best performing model, kernel regression, predicting League flexibility. The features are from top to bottom in the order of highest to lowest feature importance to the model. The X-axis indicates the directionality a feature has on the model output.}
    \label{fig:league_shap}
\end{figure*}

\begin{figure*}[!tb]
    \centering
    \includegraphics[width=\textwidth,keepaspectratio]{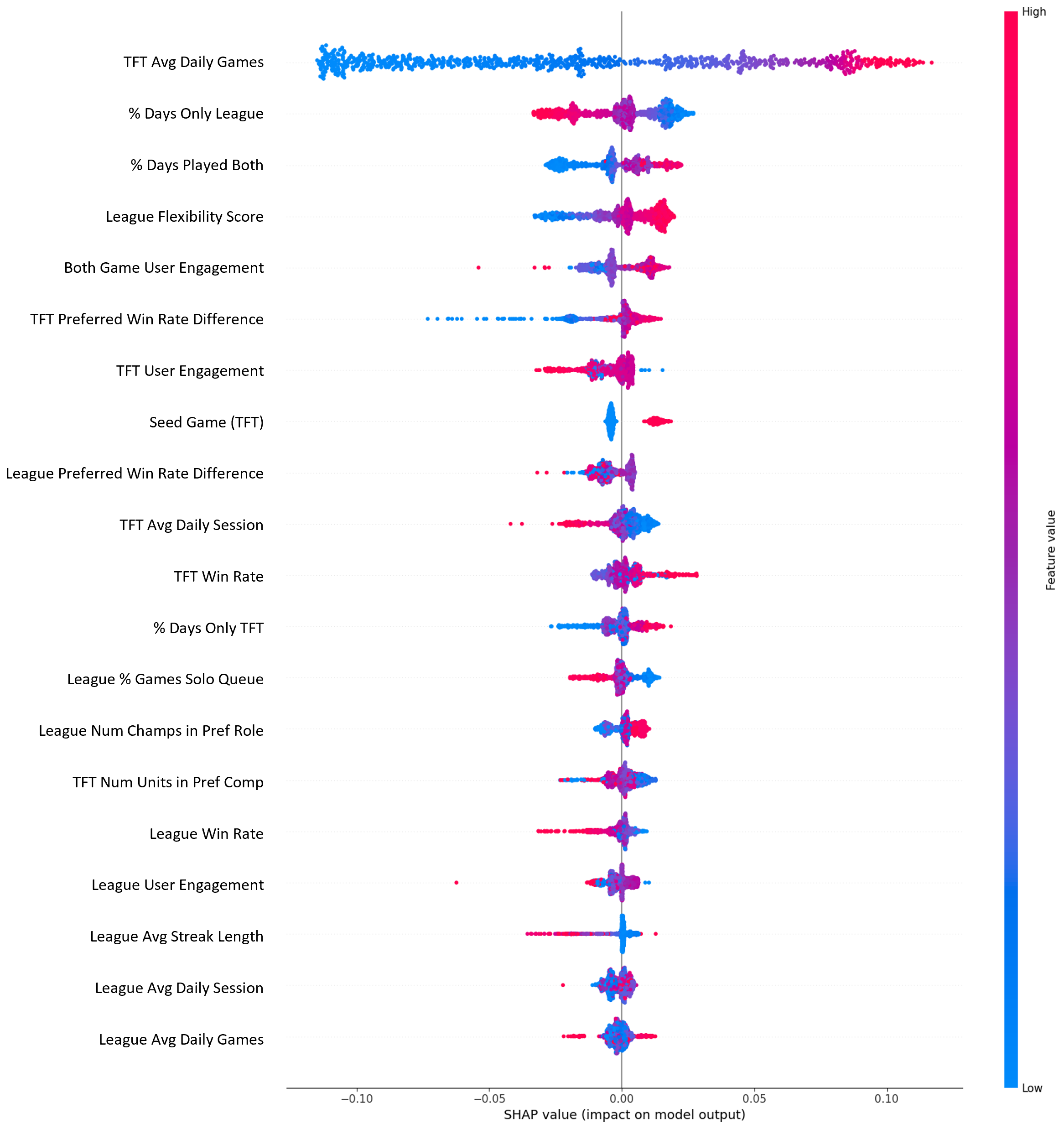}
    \caption{SHAP visualization with top 20 features for our best performing model, kernel regression, predicting TFT flexibility. The features are from top to bottom in the order of highest to lowest feature importance to the model. The X-axis indicates the directionality a feature has on the model output.}
    \label{fig:tft_shap}
\end{figure*}

\subsection{Flexibility Model Analysis}
\textbf{Interpreting SHAP Visualizations:} We used SHAP (SHapley Additive exPlanations) to visualize the impact individual features have on the performance of our models, effectively quantifying feature contribution to a model's predictive power. 
In our SHAP plots (Figures~\ref{fig:league_shap} and~\ref{fig:tft_shap}), each dot represents an individual instance (player prediction). 
The x-axis shows the magnitude and directionality of a feature's contribution to the dependent variable, while the color of each point corresponds to the feature value itself.
This color ranges from low values in blue to high values in red/pink, which are relative to the distribution across all players included in the SHAP visualization. 
We included the features with the 20 highest SHAP values in our plots sorted from highest to lowest; higher SHAP values indicate that a feature has a greater impact on a model's prediction~\cite{lundberg2017shap}.

\subsubsection{League Flexibility}
\label{sec:league_shap}
Figure~\ref{fig:league_shap} shows the SHAP visualization for the top 20 features that contribute to the kernel regression model predicting League flexibility. 
We found that a larger win rate difference (whether positive or negative) between a player's preferred play style and their overall win rate is associated with higher flexibility. 
Those who experience performance imbalance in any direction between play styles are more likely to explore other play styles. 

Players with a wider variety of options within their preferred play style are less likely to engage with other functional play styles. 
This insinuates that players may be satisfied with the opportunity to diversify within their preferred play style, decreasing the need to branch out and thus decreasing play style flexibility. 

Surprisingly, players who play a higher percentage of competitive games tend to exhibit greater flexibility, while our earlier finding that competitive success corresponds with decreased flexibility (specialization) is corroborated. 
This seemingly counterintuitive finding, especially for elite players, supports H3 and may imply that higher-skilled players have achieved a baseline of success and are more comfortable experimenting to broaden their skill sets by adapting to a wider variety of roles -- potentially to meet team needs. 
This is further supported by work from ~\citet{kim2017collective}, who found that collective intelligence, which assesses a team's capacity to work effectively across a broad set of tasks, predicts competitive success. 

The most notable takeaway from the SHAP visualization -- and a key finding for H2 -- is that TFT flexibility scores are the second most important (second highest SHAP value) predictor of League flexibility, even more so than a player's competitive success in League. 
This supports H2: players tend to maintain similar flexibility patterns across both League and TFT, suggesting that cross-game behavioral preferences persist despite competitive performance in League benefiting from specialization as demonstrated in H1. 

\subsubsection{TFT Flexibility}

Figure~\ref{fig:tft_shap} shows the SHAP visualization for the 20 top features that contribute to the kernel regression model predicting TFT flexibility. 
In TFT, the average number of daily TFT games played is the most influential predictor, suggesting that increased engagement leads to greater TFT flexibility. 
There is some inherent variance in TFT with intentional randomness built into the game design, which may encourage players to adapt their strategies more frequently over longer periods of play on the same day. 
However, increased temporal engagement in TFT decreases flexibility in players, implying that increasing temporal play frequency may lead to reliance on familiar compositions.  

Parallel to our findings with League (detailed in Section~\ref{sec:league_shap}) and directly addressing H2, we also found evidence that cross-game flexibility (League flexibility) was a more significant predictor of TFT flexibility than competitive success in TFT itself. 
League flexibility is the fourth most important feature, while competitive success in TFT ranks 11th.
This further reinforces our earlier findings that flexibility in one game positively correlates to flexibility in the other, demonstrating that many players maintain consistent behavioral patterns across games with conflicting performance incentives (H2), even though each game's structure rewards different approaches (H1).

\section{Discussion}
Our study investigates how game structure and individual agency both shape player flexibility in competitive environments, using Riot Games' League and TFT as case studies. 
By studying the same players across both games, we are able to begin disentangling and isolating the influence of game mechanics and structure from individual decision-making. 
This allows us to address the common challenge of studying these factors without introducing additional confounders such as selection bias into the agency-structure debate.
We conceptualized flexibility as the breadth of a player's engagement in different functional play styles (in League) or compositions (in TFT). 

Our results revealed that H1 was supported: League tends to reward play style specialization, evidenced by a negative correlation between flexibility and competitive success. 
TFT, on the other hand, rewards adaptability, with higher flexibility scores linked to improved performance. 
Despite these differing competitive incentives, H2 was also corroborated as players consistently exhibit similar behavioral patterns -- those who are flexible in one game are likely to be flexible in the other and vice versa. 
We also observed that elite players exhibited some adaptation to each game's optimal strategy while maintaining their baseline dispositional tendencies, demonstrating partial rather than complete behavioral change, supporting H3.
These findings suggests that underlying individual agency plays a key role in behavior even in contrasting structural contexts. 

In the following subsections, we first interpret these findings through Structuration Theory's lens of agency and structure, demonstrating how our empirical results support the duality of structure and agency. 
We then examine the psychological mechanisms underlying these patterns, drawing on the competence motivation and personality theory research to explain why players maintain dispositional preferences despite opposing structural incentives.
Finally, we translate these theoretical insights into practical implications for game designers and HCI researchers, before expanding on how our findings contribute to broader frameworks of behavioral persistence and adaptation.

\subsection{Revisiting Agency versus Structure}
A key contribution of our work provides empirical evidence in support of Giddens' Structuration Theory at scale~\cite{Giddens1979agency}. 
Unlike prior studies that have relied on smaller-scale surveys or self-reports, our large-scale natural experiment emphasizes the persistence of individual predispositions (agency) despite structural shifts. 

Structuration Theory posits that structure and agency exist in a recursive relationship: structures enable and constrain individual action, while repeated actions reproduce or transform those structures~\cite{Giddens1979agency,stones2017structuration}. 
In our study, players' flexibility patterns represent routinized practices developed through repeated gameplay. 
When entering a new game instance, players bring these established routines with them, and these preferences persist because they operate as practical consciousness, or tacit knowledge enacted automatically rather than through deliberate strategic calculation~\cite{Giddens1979agency,giddens1984constitution}.
This explains why structural incentives alone are insufficient to override established behavioral patterns: changing deeply routinized practices requires significant cognitive effort and deliberate intervention.

However, Structuration Theory also accounts for the partial adaptation we observe among elite players through reflexive monitoring -- or an actor's capacity to consciously evaluate their practices when needed~\cite{Giddens1979agency}.
Elite players, who are likely highly invested in competitive success, engage in more deliberate reflection about their strategic choices, allowing them to partially modify their flexibility patterns in response to structural incentives while retaining baseline dispositional tendencies. 
This demonstrates the duality of structure: structure exerts pressure on agency, and agency adapts somewhat, but competitive incentives (structure) are not enough; player behavior (agency) is resistant to change even in the highest levels of competitive play despite environments designed to reward opposing behavior.

These mechanisms explain our empirical findings. 
The support for H2 shows that flexibility patterns operate as routinized practices at the level of practical consciousness. 
The partial support for H1 among elite players demonstrates reflexive monitoring in action (H3), as elite players consciously evaluated their practices and partially adapted; however, the duality of structure meant that this adaptation was constrained by established dispositions. 
Even players with the strongest motivation and clearest competence incentives could not fully override their baseline flexibility tendencies.

While Structuration Theory provides a framework for understanding this agency-structure duality, SDT offers a complementary lens for understanding the psychological mechanisms underlying this behavioral persistence. 
This helps to answer the question of why players maintain certain preferences even when structural incentives suggest otherwise.

\subsection{Competence Motivation and Dispositional Persistence as Psychological Mechanisms}
Our findings reveal the interplay between the competence motivation and dispositional personality traits in shaping behavior.
The persistence of flexibility patterns across League and TFT despite their opposing performance-based incentives demonstrates that dispositional behavioral preferences -- stable personality-level tendencies toward variety-seeking or specialization -- persist even when structural incentives reward different behaviors.
This is particularly striking among elite players, who presumably have both the skill and motivation to optimize for competitive success and yet still maintain a level of cross-game behavioral consistency, suggesting that dispositional preferences constrain even motivated adaptation.

However, we also observed that elite players do shift their flexibility somewhat toward what each game rewards, indicating that the competence motivation drives partial adaptation. 
Elite players' greater tendency to optimize their behavior for competitive success demonstrates that when the competence satisfaction is highly valued, players engage in reflexive monitoring to consciously modify behavior while maintaining baseline dispositional tendencies.
This aligns with research showing that extrinsic motivations can be internalized when they serve personally important goals~\cite{ryan_self-determination_2000}, with competitive achievement representing such a goal for elite players.

Together, these theoretical perspectives provide complementary explanations for our empirical findings.
The competence motivation explains why elite players show partial adaptation (H1, H3), while personality theory research explains why dispositional preferences persist across contexts (H2).
Structuration Theory synthesizes these mechanisms through the duality of structure.
The competence motivation activates reflexive monitoring (enabling partial adaptation), but routinized dispositional practices constrain complete behavioral change (ensuring persistence).
These insights carry important practical implications for how systems should be designed to balance structural incentives with individual agency.

\subsection{Implications for Game Design and HCI}
Our results carry several practical implications for both game developers and HCI researchers. 
While structural incentives (e.g., rewards for specialization or flexibility) are crucial in shaping player behavior in competitive environments, players also bring pre-existing preferences that resist complete structural determination. 
Both SDT's competence motivation and Structuration Theory predicted this pattern: the competence motivation predicts that highly motivated players will adapt to optimize performance, while personality theory research and Structuration Theory's concept of routinized practices predict that dispositional preferences will constrain complete adaptation.
In practice, this means designers can nudge behavior, particularly among elite players who may be more motivated by the pursuit of competence, but cannot fully expect to override dispositional tendencies toward diversity or specialization in play style.

These findings extend beyond the specific context of League and TFT to highlight fundamental aspects of player identity and behavior across digital environments. 
The stability of flexibility preferences across games with opposing incentives demonstrates that behavioral dispositions constitute a core component of how users engage with systems, not merely context-dependent adaptations. 
This has implications for any domain where users navigate multiple platforms with different design philosophies, from social media to productivity tools to educational systems. 
Designers should recognize that users carry behavioral signatures across contexts, and that cross-platform behavioral patterns may be more predictive of user preferences than within-platform metrics alone.

For game designers, understanding the interplay between structural constraints and individual agency can inform the magnitude of balance updates, competitive incentives and progression rewards. 
This can include monetization or character suggestion tactics that cater towards user preferences based on their measured flexibility or rigidity. 
For researchers, the persistence of cross-game behavioral patterns underscores the challenge of precipitating change to ingrained behaviors -- this suggests that substantial structural or environmental changes may be required to induce lasting change. 
Understanding how players adapt their behaviors across multiple games and structural incentives can also help guide the development of more fluid, personalized and engaging cross-platform experiences. 

Beyond these immediate practical applications, our findings also contribute to broader theoretical questions about behavioral persistence and adaptation across contexts, which we explore in the following subsection.

\subsection{Expanding the Agency-Structure Framework}
Our study reinforces the notion that both agency and structure play an essential role in shaping player behavior, although the relative importance of either of these factors can be highly dependent on the context. 
We found that elite players are more likely to optimize their behavior to maximize performance-based success in highly competitive settings, but that this behavior is still constrained by their baseline tendencies. 
This observation aligns with prior literature on habit formation and behavior change, which suggests that while external structural factors (like game design and competitive incentive) is critical to habit formation, an individual's preferences, experience and intention ultimately dictate whether these changes are adopted and sustained~\cite{lally2013promoting,ersche2017habit}.

This study also raises the importance of temporal dynamics of player flexibility. 
Current research often treats player behavior as static, but our results suggest that player flexibility tendencies can actually evolve over time with experience or exposure to different environments. 
Although our measure of skill was limited to a player's initial rank, future research should investigate long-term behavioral trajectories to better understand lasting impacts of game design and structure on player agency and behavior. 

\subsection{Limitations and Future Work}
Our study offers new insights and corroborates the existing theory; however, our study has several limitations. 

First, our study examines two games from the same developer, Riot Games. 
While our findings of cross-game behavioral consistency align with personality research demonstrating substantial stability in dispositional preferences~\cite{roberts2000rank,fleeson2001toward}, we acknowledge that our study cannot definitively rule out the possibility that players engaging with both League and TFT represent a specific subset predisposed toward behavioral consistency. 
Players who choose to play both games may differ systematically from those who play only one, and it remains possible that broader player populations exhibit greater behavioral plasticity across gaming contexts than our sample suggests. 
Future work tracking players across more diverse game selections and ecosystems would help address this concern.

The shared ecosystem was essential for tracking individual users across games to reduce self-selection bias~\cite{Heckman2018bias,bethlehem2010bias}; however, this does limit external validity. 
Players who engage with both League and TFT may represent a specific subset with predispositions toward behavioral consistency, and both games share design philosophies that may not be representative of the broader gaming industry.
Future work should extend this analysis to games from different developers and ecosystems while maintaining individual-level tracking through, for example, platform-level account systems. 
Examining behavioral consistency across more diverse genres (e.g., first-person shooters) would further test whether flexibility operates as a stable dispositional trait across fundamentally different gameplay paradigms.

Our definition of flexibility relies on end-game snapshots of functional play style choices which may not capture the nuances of dynamic in-game decision making that may also contribute to a user's flexibility.
We operationalize flexibility as the culmination of strategic decisions reflected in end-game outcomes: in League, the champion selected after navigating pick/ban phases; in TFT, the final board composition is constructed after dynamic unit acquisition and positioning. 
These end-game snapshots capture different decision-making processes. 
League's flexibility occurs primarily before gameplay through champion selection, while TFT's can involve continuous adaptation during gameplay. 
While these operationalizations capture flexibility at different temporal stages of play and do not reflect identical constructs, both represent the final strategic choice a player commits to, and our analytical focus on within-game comparisons (elite vs. non-elite) and within-individual cross-game patterns demonstrates that behavioral dispositions influence these choices beyond what game mechanics alone would predict.
Due to the nature of user selection, access and data availability, we were also unable to survey users for their personalities or motivations. 
Although we defined flexibility more simply, future work should expand this definition to incorporate more granular in-game data (e.g., item build paths in League, units purchased and sold in TFT) and user surveys to better capture the full scope of decision-making processes. 

Our dataset only contains players from the North American servers due to API rate limits, and this may cause our findings not to generalize to player bases from other regions. 
There is evidence that cultural factors influence behavior and values both in and outside of games~\cite{triandis1989cultural,miller1984culture,karahanna2005levels,kordyaka2023exploring}, so we hope to include data from different regions to explore cultural effects in future work.

The API's rate limit also informed our sampling strategy, resulting in an oversampling of elite players compared to their non-elite counterparts. 
The binary classification of elite versus non-elite players aggregates players from Iron through Diamond into a single category, which could obscure skill-related variations in behavioral persistence within non-elite ranks; however, disentangling these effects falls outside of this paper's primary focus on comparing high-competence individuals (elite players) with the broader competitive population. 
Future work with more granular analysis across all rank tiers could reveal, for instance, whether behavioral persistence varies continuously with skill level or whether threshold effects emerge at other ranks.

Finally, the primary focus of this paper has been on a user's flexibility or adaptability; future studies could extend this work to examine player retention, social network dynamics or the impact of casual versus competitive modes on behavioral consistency. 

\section{Conclusion}
Our study provides new insights into the role that agency and structure have in shaping player behavior across different competitive gaming environments. 
We focus on a player's flexibility, or the range of play styles or compositions that a user is willing to play. 
While game design and structure influence player flexibility -- rewarding specialization in League of Legends and adaptability in Teamfight Tactics for performance-based success -- our findings show that individual tendencies still persist across these two titles. 
Performance-based structural incentives encourage opposing behaviors; however, players still maintain relatively consistent patterns of flexibility, demonstrating that personal agency plays a larger role in shaping player behavior. 
This has practical implications for game designers and researchers, who should account for both structural incentive and diversity of player preference when designing virtual environment mechanics and progression systems aimed at encouraging certain behaviors. 

Our work contributes to the broader theoretical debate on agency versus structure, providing evidence that supports Giddens' Structuration Theory~\cite{Giddens1979agency}. 
The interplay between structure and agency is complex, but can help developers and researchers construct more engaging and personalized intervention systems aimed at behavioral change that balance external rewards with individual agency.

\section{Ethics}
All data we collected are publicly accessible through the Riot API. In this work, we report our findings and analyses in aggregate to preserve user anonymity and privacy. Data are available upon reasonable request. 
\begin{acks}
The authors gratefully acknowledge the HUMANS lab and the Annenberg Game Lab for their insightful feedback and discussions on this work. 
\end{acks}
\bibliographystyle{ACM-Reference-Format}
\bibliography{sample-base}

\appendix
\section*{Appendix}
\section{Feature Descriptions}
\label{sec:feature_descriptions}
\subsection{General Variables}

\textbf{\% Days Played Both Games}: Percentage of days a user plays at least one game of League and TFT on the same day. \\
\textbf{League Streak Greater}: Indicates whether the average daily League streak is greater than the average TFT streak. Streaks are calculated by the maximum number of consecutive games of a particular title on a given day. \\
\textbf{Same Role Preference}: Indicates if a user's role preference in League matches the League roles of the units chosen in TFT. While TFT may change the functionality of units borrowed from the League IP, the unit name and themes generally remain the same. This is a boolean that is set to true if a user's most played units in TFT have the same functional play style as the user's preferred functional play style in League. \\
\textbf{Seed Skill (Non-Elite)}: A dummy variable that reflects a user's initial seed skill category: elite or non-elite. \\
\textbf{Seed Game (TFT)}: A dummy variable that reflects a user's seed game (leaderboard from which a user was retrieved). \\

\subsection{TFT-Specific Variables}
\textbf{Win Rate}: Solo Queue win rate for TFT. A user is considered to have won if they place in the top four placements in a game. \\
\textbf{Avg Daily Session}: The average number of TFT Solo Queue games a user plays per day on days that a user is playing TFT. \\
\textbf{Avg Daily Games}: The average number of TFT Solo Queue games a user plays per day. This is the total number of Solo Queue TFT games divided by the total number of days from the user's first appearance in the dataset to the present.\\
\textbf{\% Games Played Solo Queue}: The percentage of all TFT games played that are Solo Queue games. \\
\textbf{Avg Flex Score}: Flexibility scores are calculated for each user on a per-set basis based on the Solo Queue games played. These flexibility scores are then averaged to create an average flexibility score. \\
\textbf{Composition Preference Win Rate Difference}: The difference between a user's win rate while playing their preferred composition in TFT Solo Queue, versus the user's general Solo Queue win rate in TFT. \\
\textbf{Num Possible Units in Comp}: The number of total possible playable units in a user's preferred TFT composition. This accounts for compositions that might have more units available to play than others. \\
\textbf{User engagement}: This is the linear regression coefficient of a rolling window of the number of TFT Solo Queue games played in 7 days (a week), to capture user engagement. \\
\textbf{\% Days Played Only TFT}: Percentage of days a user plays at least one game of TFT, and no games of League. \\
\textbf{Avg Daily Streak Length}: Average number of consecutive TFT games played on a single day. \\

\subsection{League-Specific Variables}
\textbf{Win Rate}: Solo Queue win rate for League.  \\
\textbf{Avg Daily Session}: The average number of League Solo Queue games a user plays per day on days that a user is playing League. \\
\textbf{Avg Daily Games}: The average number of League Solo Queue games a user plays per day. This is the total number of Solo Queue League games divided by the total number of days from the user's first appearance in the dataset to the present.\\
\textbf{\% Games Played Solo Queue}: The percentage of all League games played that are Solo Queue games. \\
\textbf{Flex Score}: A user's flexibility score, calculated over their Solo Queue League of Legends games. 
\textbf{Role Preference Win Rate Difference}: The difference between a user's win rate while playing their preferred functional play style in League Solo Queue, versus the user's general Solo Queue win rate in League. \\
\textbf{Num Units in Role}: The number of total possible playable champions in a user's preferred League functional play style. This accounts for roles that might have more champions available to play than others. \\
\textbf{User engagement}: This is the linear regression coefficient of a rolling window of the number of League Solo Queue games played in 7 days (a week), to capture user engagement. \\
\textbf{\% Days Played Only League}: Percentage of days a user plays at least one game of League, and no games of TFT.  \\
\textbf{Avg Daily Streak Length}: Average number of consecutive League games played on a single day. \\


\section{League flexibility vs. TFT flexibility}
\label{sec:league_tft_flex_tab}
Table~\ref{tab:flex50_flipped} summarizes 50th-percentile flexibility scores by seed and skill group.

\begin{table}[h]
\centering
\begin{tabular}{l|c|c}
\textbf{Group (Seed, Skill)} & \textbf{50\%-tile League Flex} & \textbf{50\%-tile TFT Flex} \\ \hline
League, Elite      & 0.603527 & 0.708921 \\
League, Non-Elite  & 0.646259 & 0.716369 \\
TFT, Elite         & 0.871145 & 0.627450 \\
TFT, Non-Elite     & 0.808575 & 0.661429 \\
\end{tabular}
\captionof{table}{50th-percentile flexibility scores disaggregated by user seed and skill group.}
\label{tab:flex50_flipped}
\end{table} 

\section{Optimal Hyperparameters}
\label{sec:appendix_hyperparams}
\subsection{GBM Hyperparameters}
\begin{table}[h]
\centering
\begin{tabular}{c|c|c}
\textbf{Metric} & \textbf{League Flex} & \textbf{TFT Flex} \\ \hline
\textbf{Learning Rate} & 0.01 & 0.05 \\ 
\textbf{Max Depth} & 5 & 5 \\
\textbf{Max Features} & None & sqrt \\ 
\textbf{Min Samples Leaf} & 1 & 1 \\ 
\textbf{Min Samples Split} & 10 & 5 \\ 
\textbf{Number of Estimators} & 300 & 200 \\ 
\textbf{Subsample} & 0.8 & 0.8 \\ 
\textbf{MAE} & 0.1499 & 0.0840 \\ 
\textbf{MSE} & 0.0341 & 0.0121 \\ 
\textbf{RMSE} & 0.1847 & 0.1102 \\ 
\textbf{R\textsuperscript{2}} & 0.2093 & 0.4438 \\ 
\end{tabular}
\captionof{table}{Best performing GBM hyperparameters and performance metrics for predicting TFT and League flexibility after hyperparameter tuning.}
\label{tab:gbm_hyperparameter}
\end{table}
\newpage

\subsection{Kernel Regression Hyperparameters}
\begin{table}[h]
\centering
\begin{tabular}{c|c|c}
\textbf{Metric} & \textbf{League Flex} & \textbf{TFT Flex} \\ \hline
\textbf{Alpha} & 0.57 & 0.4 \\ 
\textbf{Kernel} & laplacian & laplacian \\ 
\textbf{Gamma} & 0.0098 & 0.01 \\ 
\textbf{Test Set MSE} & 0.03501 & 0.01161 \\ 
\textbf{Test Set R\textsuperscript{2}} & 0.2235 & 0.45513 \\ 
\end{tabular}
\captionof{table}{Best performing kernel regression hyperparameters and performance metrics for predicting TFT and League flexibility after hyperparameter tuning.}
\label{tab:kernelreg_hyperparameter}
\end{table}
\subsection{Neural Network Hyperparameters}
\begin{table}[h]
\centering
\begin{tabular}{c|c|c}
\textbf{Metric} & \textbf{League Flex} & \textbf{TFT Flex} \\ \hline
\textbf{Layers} & (128, 64, 32) & (512, 256, 128, 64, 32) \\ 
\textbf{Activation} & sigmoid & sigmoid \\ 
\textbf{Optimizer} & sgd\_momentum & nadam \\
\textbf{Batch Size} & 17 & 32 \\
\textbf{Epochs} & 75 & 75 \\
\textbf{Dropout Rate} & - & 0.2 \\ 
\textbf{Weight Regularization} & - & 0.001 \\ 
\textbf{Test Set MSE} & 0.03537 & 0.011871 \\
\textbf{Test Set R\textsuperscript{2}} & 0.15261 & 0.437512 \\ 
\end{tabular}
\captionof{table}{Best performing neural network hyperparameters and performance metrics for predicting League and TFT flexibility after hyperparameter tuning.}
\label{tab:nn_hyperparameters}
\end{table}

\end{document}